\documentclass{jpp}
\usepackage{graphicx}
\usepackage{epstopdf, epsfig}
\usepackage{hyperref}
\usepackage{amsmath,amsfonts,amssymb}
\usepackage{subfigure}

\usepackage[normalem]{ulem}
\usepackage{color} 

\newcommand{\sthet}{\sin\theta} 
\newcommand{\er}{\vec{e}_r}	
\newcommand{\ethet}{\vec{e}_\theta}
\newcommand{\ephi}{\vec{e}_\varphi}
\newcommand{\ppot}{{\cal P}}
\newcommand{\vB}{\vec{B}}
\newcommand{\vnabla}{\vec{\nabla}}
\newcommand{\mubar}{\bar \mu}
\newcommand{\Apot}{{\cal A}}

\shorttitle{Force-free twisted magnetosphere}
\shortauthor{G. Voisin}

\title{Twisted magnetar magnetospheres: a class of semi-analytical force-free non-rotating solutions}

\author{G. Voisin\aff{1}\corresp{\email{guillaume.voisin@obspm.fr}}}

\affiliation{\aff{1} LUX, Observatoire de Paris, Université PSL, Sorbonne Université, CNRS, 92190 Meudon, France}

\begin{document}

\maketitle

\begin{abstract}
Magnetospheric twists, that is magnetospheres with a toroidal component, are under scrutiny due to the key role the twist is believed to play in the behaviour of neutron stars. Notably, its dissipation is believed to power magnetar activity, and is an important element of the evolution of these stars.
We exhibit a new class of twisted axi-symmetric force-free  magnetospheric solutions. 
We solve the Grad-Shafranov equation by introducing an ansatz akin to a multipolar expansion. We obtain a hierarchical system of ordinary differential equations where lower-order multipoles source the higher-order ones. We show that analytical approximations can be obtained, and that in general solutions can be numerically computed using standard ODE solvers.
We obtain a class of solutions with a great flexibility in initial conditions, and show that a subset of these asymptotically tend to vacuum. The twist is not confined to a subset of field lines. The solutions are symmetric about the equator, with a toroidal component that can be reversed. This symmetry is supported by an equatorial current sheet. We provide a first-order approximation of a particular solution that consists in the superposition of a vacuum dipole and a toroidal magnetic field sourced by the dipole, where the toroidal  component decays as $1/r^4$.  As an example of strongly multipolar solution, we also exhibit cases with an additional octupole component.
\end{abstract}


\section{\label{sec:intro}Introduction}

Twisted magnetospheres of neutron stars are being investigated for the role that the twist plays in the dynamics and radiation of these stars, particularly in the case of magnetars \citep{thompson_electrodynamics_2002}. By twist, it is meant that the magnetic field has a toroidal component that is supported by magnetospheric currents. Among other things, twists can affect the opacity of the magnetosphere \citep[e.g.][]{vigano_force-free_2011} because of the strong currents and particle densities they require, or affect the spin-down rate of the star by modifying the polar cap structure \citep{glampedakis_inside-out_2014}. Untwisting is proposed to be at the origin of at least some of the stellar radiation \citep[e.g.][]{beloborodov_untwisting_2009, vigano_force-free_2011}. Twists represent an energy reservoir which can provide the energy dissipated into giant magnetar flares \citep{mahlmann_instability_2019}. Recent work has been dedicated to studying the dynamics of untwisting in 3D configurations, introducing increasingly complex configurations \citep[e.g.]{ mahlmann_three-dimensional_2023, mahlmann_instability_2019, carrasco_triggering_2019}. The literature mostly focuses on dipolar boundary conditions at the stellar surface, and the case of multipolar configurations has had relatively little attention.

The equation describing an aligned non-rotating force-free axisymmetric magnetosphere is the Grad-Shafranov equation, and the so-called pulsar equation \citep{michel_rotating_1973} can be seen as its extension to rotating, relativistic, magnetospheres. A few semi-analytical solutions exist for non-rotating twisted magnetospheres \citep{wolfson_shear-induced_1995, thompson_electrodynamics_2002}, and see \citet{vigano_force-free_2011} for a review. In particular, self-similar solutions have the property that both poloidal and toroidal components of the magnetic field decay with the same power-law of the radius. These solutions therefore do not asymptotically tend to vacuum. Self-similar solutions allow for multipolar-like fields \citep{pavan_topology_2009}, but a single multipole is possible at a given time because the non-linearity of equations precludes linear combinations. Much work has recently been done on numerical solutions where the twist is confined, by construction, to a subset of field lines. This presents the advantage of allowing for vacuum fields outside of this region \citep[e.g.][and references therein]{akgun_force-free_2016, ntotsikas_twisted_2024}. Some of this work has been focused on the matching of surface boundary conditions with internal solutions in order to study coupled equilibrium and evolution \citep{vigano_force-free_2011, fujisawa_magnetic_2014, pili_general_2015, glampedakis_inside-out_2014, akgun_long-term_2017, uryu_equilibriums_2023}. In \citet{parfrey_dynamics_2013}, it is shown that beyond a certain critical twist the force-free magnetosphere becomes tearing-mode unstable, forms a current sheet that dissipates the energy through magnetic reconnection. The existence of a critical twist beyond which no stationary solution is possible has been recognised by many authors \citep[e.g.][and references therein]{akgun_force-free_2016}.

In this work we present a class of semi-analytical solutions to the non-rotating Grad-Shafranov equation akin to a multipolar expansion of the magnetic field. It allows for the choice of an arbitrary set of multipoles and can asymptotically tend to a vacuum dipole. In Sec. \ref{sec:math} we briefly review the mathematical framework, in Sec. \ref{sec:solution} we present the class of solutions, in Sec. \ref{sec:application} we present a couple of examples of solutions, and we conclude in Sec. \ref{sec:conclusion}.

\section{Mathematical framework} \label{sec:math}
For a non-rotating star, the force-free condition $\vec{j}\times \vec{B} = 0$ \citep[e.g.][]{gruzinov_force-free_2006} leads to
\begin{equation}
\label{eq:ff}
\vnabla \times \vB = \alpha \vB,
\end{equation}
where $\alpha$ is a function such that $\vec{j} = \alpha \vec{B}$ is the current density.  

Taking the divergence of Eq. \eqref{eq:ff},
\begin{equation}
\label{eq:fldep}
\vB\cdot\vnabla \alpha = 0,
\end{equation}
such that $\alpha$ is constant along a field line. 

In the following we seek cylindrically-symmetric solutions, such that $\partial_\varphi = 0$ where $\varphi$ is the angle around the symmetry axis. The system is therefore effectively two-dimensional, and in a system of coordinates where one of them labels magnetic field lines, Eq. \eqref{eq:fldep} implies that $\alpha$ depends on that single coordinate. 
Because of axial symmetry the poloidal part of the magnetic field is itself divergence-free. This allows one to express it through a single Euler potential \citep[e.g.][]{stern_euler_1970} $\ppot$,
\begin{equation}
\label{eq:Bp}
\vB_{\rm p } = \vnabla \ppot \times \vnabla \varphi =  \frac{\vnabla \ppot \times \ephi}{r\sthet} = \frac{r^{-1} \partial_\theta \ppot \er - \partial_r \ppot \ethet}{ r\sthet},
\end{equation}
where $\ppot$ labels magnetic field lines \footnote{Strictly speaking poloidal field lines, but these map to general field lines through axial symmetry.}. It follows that one can write $\alpha = \alpha(\ppot)$. The magnetic field reads $ \vB= \vB_{\rm p } + B_{\varphi} \ephi$.  

Expending Eq. \eqref{eq:ff} in spherical coordinates, we get
\begin{equation}
\label{eq:ffexp}
\left(\begin{matrix}
\frac{1}{r\sthet} \partial_\theta \left(\sthet B_{\varphi}\right) \\
-\frac{1}{r}\partial_r\left(r B_{\varphi}\right) \\
\frac{1}{r} \left(\partial_r \left(r B_{\theta}\right) - \partial_\theta B_{r}\right)
\end{matrix}\right) = \alpha \left(\begin{matrix}
B_{r} \\
B_{\theta} \\
B_{\varphi}
\end{matrix}\right).
\end{equation}

From Eq. \eqref{eq:Bp} we see that the poloidal part of the left-hand side of Eq. \eqref{eq:ffexp} can be written 
\begin{eqnarray}
\label{eq:ffexpdetail}
\frac{1}{r\sthet} \partial_\theta \left(\sthet B_{\varphi}\right) & = & \alpha \frac{\partial_\theta \ppot }{r^2\sthet}, \\
\frac{1}{r}\partial_r\left(r B_{\varphi}\right) & = & \alpha \frac{\partial_r \ppot}{ r\sthet}. \nonumber
\end{eqnarray}
Since we have seen that $\alpha$ is a function of $\ppot$, we can introduce the primitive $A = \int \alpha \mathrm{d} \ppot$, and see from Eq. \eqref{eq:ffexpdetail} that 
\begin{equation}
\label{eq:A}
A = r\sthet B_{\varphi}. 
\end{equation}

Inserting Eq. \eqref{eq:Bp} into the third component of the force-free equation, Eq. \eqref{eq:ffexp}, we obtain the Grad-Shafranov equation
\begin{equation}
\label{eq:gseq}
-\partial_r^2\ppot - \frac{1-\mu^2}{r^2} \partial_\mu^2\ppot = \alpha(\ppot) A(\ppot),
\end{equation}
where $\mu \equiv \cos\theta$. 
Boundary conditions must be such that $\ppot(\mu=\pm 1) = 0$ which ensures that the field line going out of either side of the symmetry axis is the same and avoids field-line crossings \citep{wolfson_shear-induced_1995}.

More generally, the boundary conditions of this equation are set by specifying $\ppot$ on the stellar surface, and at infinity. Below and in Sec. \ref{sec:solution} we present solutions that are decomposed on a functional basis such that asymptotically $\ppot \propto 1/r$ and therefore the external boundary condition is automatically satisfied. The treatment of the inner boundary condition then depends on the class of solutions considered.

It is instructive to look at the example of self-similar solutions to Eq. \eqref{eq:gseq} \citet{lynden-bell_self-similar_1994,low_modeling_1990,wolfson_shear-induced_1995, thompson_electrodynamics_2002}. It consists in using the ansatz 
\begin{equation}
\label{eq:selfsimsol}
\ppot = F(\mu)/r^p \;\; ; \;\; \alpha(\ppot) = c \ppot^{1/p},
\end{equation} 
where $c$ is a constant, $p \geq 1$ is a constant index, and $F$ is a function to be determined. An ordinary differential equation for $F$ is obtained by inserting Eq. \eqref{eq:selfsimsol} into \eqref{eq:gseq}. This equation is non-linear and can be integrated numerically. In order to proceed, one must specify the conditions $F(\mu=1) =\ppot(\mu= 1) = 0$ and its derivative with respect to $\mu$, $F'(\mu=1)$. The latter corresponds to the radial magnetic field at the pole, as can be seen from Eq. \eqref{eq:Bp}. However, if setting $(F(1),F'(1))$ determines the initial value problem for numerical integration, there is no guarantee that the boundary value problem $F(\mu= -1) =\ppot(\mu= -1) = 0$ will be met as well. In fact, for fixed $c$ and $p$ boundary conditions are verified for a discrete set of $F'(\mu=1)$ which one must find numerically (for example by a shooting method). Since it is physically more interesting to specify $F'(\mu=1)$, one rather varies $c$, for which a discrete spectrum of solutions exist. This can be seen as a non-linear eigenvalue problem of $c(p)$. Since Eq. \eqref{eq:gseq} with the ansatz Eq. \eqref{eq:selfsimsol} is even, $\ppot(-\mu)$ is solution if $\ppot(\mu)$ is, one can integrate only until the equator at $\mu = 0$ and replace the boundary condition at $-1$ by $\partial_\theta \ppot(\mu = 0) =0$ \citep{thompson_electrodynamics_2002}.

\section{Solutions} \label{sec:solution}

\subsection{Ansatz}
\label{sec:ansatz}
Multipolar expansions are general solutions of the vacuum electromagnetic field in spherical symmetry \citep{bonazzola_general_2015, petri_multipolar_2015}. This motivates us to posit the ansatz
\begin{equation}
\label{eq:ppot}
\ppot  =  B_0 R^2 \sum_{i=1} F_i(\mu)\frac{R^i}{r^i},
\end{equation}
where $R$ is the stellar radius, $B_0$ gives the scale of the surface magnetic field intensity, and $F_i$ are functions to be determined. 

A heuristic for the induction of the ansatz in Eq. \eqref{eq:selfsimsol} consists in analysing the dependence on $r$ on both sides of Eq. \eqref{eq:gseq}. Indeed, once one has posited the ansatz for $\ppot$, it is clear that the left-hand side is $\propto 1/r^{p+2}$. From this observation the form proposed for $\alpha$ in Eq. \eqref{eq:selfsimsol} appears as the simplest way to balance out the powers of $r$. With the current ansatz Eq. \eqref{eq:ppot}, applying this heuristics brings us to postulate the following form for $\alpha$ and its integral $A$
\begin{eqnarray}
\label{eq:alphaansatz}
\alpha(\ppot) & = & R^{-1}\sum_{i=1} (i+1) c_i \left(\frac{\ppot}{B_0R^2}\right)^i, \\
\label{eq:Aansatz}
A(\ppot) & = & B_0 R \sum_{i=1} c_i  \left(\frac{\ppot}{B_0R^2}\right)^{i+1},
\end{eqnarray}
where $c_i$ are a priori free coupling constants. 

Keeping in mind that $[\ppot] = \text{magnetic strength} \times \text{length}^2$, it follows that the $F_i$ are dimensionless. Similarly, from Eq. \eqref{eq:ff} one sees that $[\alpha] = \text{length}^{-1}$, such that the coupling constants $c_i$ are also dimensionless. 
Thereafter we use units such that $B_0=R=1$.

\subsection{Source term $\alpha A$}
\newcommand{\ui}{\underline{i}}
\newcommand{\hi}{\hat{i}}
The source term of Eq. \eqref{eq:gseq} reads 
\begin{equation}
\label{eq:sterm}
\alpha A = \sum_{i,j=1} (i+1) c_i c_j \ppot^{i+j+1} = \sum_{k = 3} \sum_{i=1}^{k-2} c_i c_{\underline{j}} (i+1) \ppot^{k},
\end{equation}
where $k = i+j+1$ and $\underline{j} = k -i-1$. 

We formulate the powers of $\ppot$ as a Laurent series in $r$, 
\begin{eqnarray}
\label{eq:ppotk}
\ppot^k & = & \sum_{i_1, ...,i_k}\frac{F_{i_1}...F_{i_k}}{r^l}  \text{ with } l  = \sum_m^k i_m \\
& = & \sum_{l=k}\frac{G_{l}^{(k)}}{r^l} \\
\end{eqnarray}
with 
\begin{eqnarray}
\label{eq:g}
G_{l}^{(k)} & = & \sum_{i_1+...+i_k = l; i_j \geq 1} F_{i_1}...F_{i_k} \\
& = &  \sum_{i_1 = 1}^{\hat{i}_1}...\sum_{i_{k-1} = 1}^{\hat{i}_{k-1}}	F_{i_1}...F_{\ui_k},
\end{eqnarray}
where we have $\ui_k = l - \sum_{j = 1}^{ k-1} i_j$ and $\hi_j = l - \sum_{m=1}^{j-1} i_m - (k-j)$.
In particular it results that $G_{k}^{(k)} $ = $ F_1^k$ and, more generally one can show the following functional dependence, 
\begin{eqnarray}
\label{eq:gdep}
G_{k+m}^{(k)} & = & f\left(F_1, ..., F_{m+1}\right).
\end{eqnarray}
This results from $\forall k, \max(\ui_k) = m+1$.

We can now decompose the source term into a Laurent series with explicit coefficients. First, inserting Eq. \eqref{eq:ppotk} into Eq. \eqref{eq:sterm}, we obtain
\begin{equation}
\alpha A =  \sum_{k = 3} \sum_{i=1}^{k-2} c_i c_{\underline{j}} (i+1) \sum_{l=k} \frac{G_{l}^{(k)}}{r^l}.
\end{equation}
Second, we swap sums such that
\begin{equation}
\label{eq:stermlaurent}
\alpha A =  \sum_{l = 3} \frac{1}{r^l}\sum_{k=3}^{l}\sum_{i=1}^{k-2} c_i c_{\underline{j}} (i+1)  G_{l}^{(k)} = \sum_{l = 3}  \frac{1}{r^l} [\alpha A]_{l},
\end{equation}
where $ [\alpha A]_{l} = \sum_{k=3}^{l}\sum_{i=1}^{k-2} c_i c_{\underline{j}} (i+1)  G_{l}^{(k)}$ denotes the angular part of the $\bigcirc(r^{-l})$ term of the function $\alpha A$.
In order to generate the subsequent orders of the source we see from Eq. \eqref{eq:stermlaurent} that 
\begin{equation}
\label{eq:alphaA}
\left[\alpha A \right]_{l+1} =  \left[\alpha A \right]_{l} (l \rightarrow l+1) + G_{l+1}^{(l+1)} \sum_{i = 1}^{l-1} c_i c_{\underline{j}} (i+1)
\end{equation}
where the arrow indicates replacement of the lower indices of the $G$ functions, and $\underline{j} = l-i$.

Here we explicit the first 5 orders of the source term,
\begin{eqnarray}
\label{eq:aA3}
\left[\alpha A \right]_3 & = & 2c_1^2 G_{3}^{(3)}, \\
\label{eq:aA4}
\left[\alpha A \right]_4 & = & 2c_1^2 G_{4}^{(3)} + 5c_1 c_2 G_{4}^{(4)}, \\
\label{eq:aA5}
\left[\alpha A \right]_5 & = & 2c_1^2 G_{5}^{(3)} + 5c_1 c_2 G_{5}^{(4)} + (6c_1 c_3 + 3c_2^2) G_{5}^{(5)}, \\
\left[\alpha A \right]_6 & = & 2c_1^2 G_{6}^{(3)} + 5c_1 c_2 G_{6}^{(4)} + (6c_1 c_3 + 3c_2^2) G_{6}^{(5)} \\
&  & + 7(c_1c_4 + c_2c_3) G_6^{(6)}, \nonumber\\
\left[\alpha A \right]_7 & = & 2c_1^2 G_{7}^{(3)} + 5c_1 c_2 G_{7}^{(4)} + (6c_1 c_3 + 3c_2^2) G_{7}^{(5)} \\
&  & + 7(c_1c_4 + c_2c_3) G_7^{(6)} + \left[8(c_1 c_5 + c_2 c_4) + 4c_3^2\right] G_{7}^{(7)}. \nonumber
\end{eqnarray}

\subsection{Hierarchy} \label{sec:hierarchy}
Inserting Eq. \eqref{eq:stermlaurent} into the Grad-Shafranov equation Eq. \eqref{eq:gseq}, and solving for each coefficient of the Laurent series in $r$, it transforms into a set of ordinary differential equations, 
\begin{equation}
\label{eq:hierarchy}
\forall i \geq 1, -i(i+1) F_i - (1-\mu^2) F_i'' = [\alpha A]_{l}\left(\{F_j\}_{j\leq i}\right),
\end{equation}
where $l=i+2$ and, by virtue of Eq. \eqref{eq:gdep}, the right-hand side only depends on $F_{j \leq i}$. As a consequence of the hierarchy of this set of equations, it can be solved iteratively up to a certain truncation order. 

The boundary conditions $\ppot(\mu =\pm 1) = 0$ translate into $\forall i\geq 1, F_i(\pm 1) = 0$. Indeed we need $\ppot(\mu =\pm 1) = 0$ for all $r$ along the symmetry axis which, given the definition Eq. \eqref{eq:ppot}, implies that all $F_i$ must individually cancel at the boundaries. In practice, we will build solutions by specifying $\{F_i'(\mu=1)\}_{i\geq1}$ (see below). 

When all coupling constants are null, $c_i=0$, then each element of Eq. \eqref{eq:hierarchy} gives the corresponding vacuum multipole (see appendix \ref{ap:vacsol}): $F_1$ is a dipole, $F_2$ a quadrupole, $F_3$ an octupole, and so on. 

If $c_1 = 0$ then all equations are linear (albeit with a potentially cumbersome source term). If the dipole order is present, that is $F_1 \neq 0$, then it is a vacuum dipole as its source term is empty. Provided that some other coupling constant is non-zero, then this dipole sources an infinity of higher-order equations. For example, if $c_2$ is active, then the source term of the equation for $F_3$ is $[\alpha A]_{5} = 3c_2^2 G_{5}^{(5)} = 3c_2^2 F_1^5$, according to Eq. \eqref{eq:aA5}. Since all higher order terms decay with radius faster that the $F_1$ term, the solution is asymptotically a vacuum dipole. More generally, for $c_1=0$ solutions asymptotically tend to the first non-zero vacuum multipole at infinity. Since equations become linear, they can in principle be solved analytically using power series. However we here mostly use numerical integration as it appears simpler for our purposes. 

On the other hand equations are highly non-linear when $c_1 \neq 0$. For example, the source term of the leading order equation for $F_1$ is $2c_1^2 F_1^3$ according to Eq. \eqref{eq:aA3}. As a result, these solutions do not asymptotically connect to vacuum.

\subsection{Boundary conditions, current sheet, and regular or anti-twist}\label{sec:evenparity}
\subsubsection{Boundary conditions}
We need to satisfy the boundary conditions given by $\forall i\geq 1, F_i(\pm 1) = 0$ (see Sec. \ref{sec:hierarchy}). In principle, one possibility would be to integrate from the north pole at $\mu=1$ with $\{F_i(\mu=1)=0\}$ until the south pole at $\mu = -1$ and adjust the $\{F_i'(\mu=1)\}$, or alternatively the constants $\{c_i\}$, such that the second boundary condition $\{F_i(\mu=-1)=0\}$ be met. This, in fact, would be the generalisation of the method used to obtain self-similar solutions, Sec. \ref{sec:math}. However, we have empirically found that this approach leads to the divergence of the $\{F_i\}$ series even in the simplest cases (e.g. Sec. \ref{sec:simpledip}).

One can see that for any solution $\{F_i(\mu)\}_{i\geq 1}$ of Eq. \eqref{eq:hierarchy} on the interval $[-1,0[$, $\{F_i(-\mu)\}_{i\geq1}$ is solution on the interval $]0,1]$. This even symmetry with respect to the equatorial plane allows us to build a solution that fulfils boundary conditions at the poles and that is defined everywhere except in that plane. To this end, one needs to integrate Eq. \eqref{eq:hierarchy} with boundary conditions $\{F_i(\mu=1)=0, F_i'(\mu=1)\}_{i\geq 1}$ in, say, the northern hemisphere ($\mu >0$), and then mirror it onto the other hemisphere. Here, $\{F_i'(\mu=1)\}$ are free parameters. This procedure comes at the price of an equatorial current sheet, as explained below, but makes the satisfaction of the boundary conditions $\forall i\geq 1, F_i(\pm 1) = 0$ straightforward (see Sec. \ref{sec:hierarchy}).

The case where $c_1$ is the only non-zero constant is particular. Indeed, if all $F_i'(\mu=1) =0$ except $F_1'$, that is the dipolar order, then the hierarchy stops at $F_1$ (i.e. $F_{i>1}= 0$) and we are in the non-linear case already studied in \citep{low_modeling_1990}. In this case, there is a discrete spectrum of values of $c_1$ (or $F_1'(\mu=1)$) for which the solution satisfies the boundary conditions and that is continuous in the equatorial plane. It is possible to add higher orders to such solution, although at the price of a discontinuity at these orders. We expand on this in Sec. \ref{sec:nvdip}.

 
\subsubsection{Regularly and anti-twisted solutions}

Two such solutions are in fact possible: for a given set of coupling constants $\{c_i\}_{i \geq 1}$ in the northern hemisphere one can choose $\{\pm c_i\}_{i \geq 1}$ in the south. Indeed, the right-hand side of Eq. \eqref{eq:hierarchy} depends only on products $c_i c_j$ such that a change of sign of all coupling constants has no effect. However, the sign of $B_\varphi$ is then reversed across the equator through Eq. \eqref{eq:A}. We shall refer to solutions with reversed $B_{\varphi}$ as anti-twisted solutions and others as regularly twisted.

\subsubsection{Solutions with a current sheet} \label{sec:curdens}
The solution is constructed as follows,
\begin{equation}
\ppot = \left\{\begin{matrix}
\ppot_{\rm n}(\mu) \text{ for } \mu > 0 \\
\ppot_{\rm s} = \ppot_{\rm n}(-\mu) \text{ for } \mu < 0
\end{matrix}\right.
\end{equation}
where $\ppot_{\rm n}$ is the potential of the northern hemisphere obtained by integration of Eq. \ref{eq:hierarchy} from $\mu = 1$ to 0, while the southern hemisphere $\ppot_{\rm s}$ is obtained by symmetry. As a result of this even symmetry, $\ppot(0^{+}) = \ppot(0^{-})$ and therefore the magnetic component normal to the equator, $B_\theta \propto \partial_{r} \ppot$, is continuous across it (see Eq. \ref{eq:Bp}). On the other hand, the radial component $B_{ r} \propto \partial_{\theta} \ppot$, tangent to the equatorial plane, is generally discontinuous since $\partial_{\theta} \ppot(0^{+}) = - \partial_{\theta} \ppot(0^{-})$. That is to say, unless $B_{ r} = 0$ as is the case for a vacuum dipole, for example. As a result, the even symmetry produces an azimuthal current sheet with surface current density 
\begin{equation}
\label{eq:sigphi}
	\sigma_{\varphi} = -2B_r(\mu=0^{+}).
\end{equation} 
For anti-twisted solutions, everything is identical except for the sign of the toroidal component $B_\varphi$ which reverses across the equator. This is supported by a radial component of the current sheet with density given  by 
\begin{equation}
\label{eq:sigr}
	\sigma_r = \left\{\begin{matrix}
		-2B_r(\mu=0^{+}) \text{ (anti-twisted)},\\
		0 \text{ (regularly twisted)}.
	\end{matrix} \right.
\end{equation}

\subsubsection{Practical consequences}
For odd-order $F_i$ functions, vacuum solutions are unaffected since $B_r = 0$ in the equatorial plane (vacuum dipole, octupole, etc...). This is because these functions are odd (see appendix \ref{apsec:parity}). For example, if $c_1=0$, $F_1$ is always a vacuum dipole which does not generate a current sheet. However, particular solutions of higher orders do not share the property $B_r = 0$ which results in a toroidal current sheet $\sigma_\varphi$ as seen above. For example, a vacuum dipole sources higher orders with equatorial discontinuities supported by the current sheet. In the case with coupling constants $c_{i\neq 2}=0$, the dipole sources octupolar and higher orders, $F_{i\geq 3}$. This case is detailed in appendix \ref{ap:F3}. As a result, the current sheet is localised in the sense that it is born by components $1/r^{i\geq 5}$. 

Even-order vacuum solutions $F_i$ are odd functions. As a consequence, even symmetry makes them become what we may call split multipoles. For example, a vacuum quadrupole, which is possible provided that $c_1=0$, becomes a split quadrupole. As before, this is supported by the toroidal component of the current sheet. Discontinuities in the particular solutions behave the same.

In order to keep the natural symmetry of the quadrupole and simultaneously satisfy boundary conditions, one would need to impose odd parity to all orders. However, the quadrupole sources higher order equations in the hierarchy. Particular solutions to these equations, contrary to vacuum ones, will not in general satisfy $B_\theta(\mu =0)=0$. As a result, discontinuities arise in the magnetic component normal to the equatorial plane, which cannot be supported by a current sheet and is unphysical.

\subsection{Convergence}
A necessary condition for convergence of the series defining the potential $\ppot$, Eq. \eqref{eq:ppot}, is that the source term $\alpha A$ defined by Eq. \eqref{eq:stermlaurent} itself converges. Indeed one expects that $F_i/r^l \sim [\alpha A]_{l}/r^l$ with $l=i+2$ from Eq. \eqref{eq:hierarchy}. We conjecture that this is a sufficient condition in general as well, without being able to demonstrate it at this point. Nonetheless, all the cases studied in Sec. \ref{sec:application} are numerically shown to be converging. The special case for $c_{i\neq 2} = 0$ is discussed below and in appendix \ref{ap:convergence}.

Since with $c_{i\neq 2} = 0$ all equations are linear beyond first order, solutions are the sum of a particular and a vacuum solution. Vacuum solutions are proportional to the initial condition  $F_i'(\mu = 1)$ (see appendix \ref{ap:F3}). Therefore, a necessary condition for convergence is that $F_i'(\mu = 1) \rightarrow_{i\rightarrow \infty} 0$. For the rest, the source term tending to zero implies the particular solutions also tending to zero.  

Concerning the $F_i$ functions, we have not had any convergence or singularity issue in all the cases that have been studied numerically. We have analytically studied the particular case of the function $F_3$ with $c_{i\neq 2}=0$, Sec. \ref{ap:F3}, shows that indeed the solution is analytical.

\section{Applications}\label{sec:application}
\newcommand{\ve}{\vec{e}}
\newcommand{\Bd}{B_{\rm d}}

In this section, we first study the simplest case that produces i) asymptotically a vacuum dipole and ii) the leading order toroidal component. Such configurations are obtained for $c_2 \neq 0$ and $c_{i\neq 2} = 0$. In all the cases presented, we present plots corresponding to the anti-twisted solutions unless otherwise mentioned. This choice is only made to limit the number of plots to a minimum. Indeed, the regularly twisted configurations can be straightforwardly deduced from the anti-twisted ones by symmetrising with respect to the equator, and setting the radial current-sheet component to zero, that is $\sigma_r =0$. 

We also discuss the case $c_{i\neq 1} = 0$, where the dipole order is not in vacuum, but rather interacts non-linearly. This case was already studied in previous literature but without the possibility of adding multipoles. Here we exhibit such example.

For practical purposes it is convenient to carry out the integration of Eq. \eqref{eq:hierarchy} as a function of $\mubar = 1 -\mu$. In the following we denote by a dot the derivative with respect to $\mubar$, and unless otherwise stated the boundary conditions are denoted $\dot F_i
= -F_i'(\mu = 1)$.

In all cases we give the maximum twist, that is the largest azimuthal shift between the two footpoints of a field line, $\Delta \varphi = \oint_{\text{field line}} \mathrm{d}\varphi$, and the magnetic helicity $H = \int \vec{\Apot} \cdot \vec{B} \;\mathrm{d} V$. We also numerically compute the magnetic energy as $E = \int B^2 \mathrm{d}V$, which we compare to the energy of the vacuum multipoles alone (i.e. $\forall i, c_i = 0$). Values for energy are given in units of $2\pi B_0^2R^3$ and values of helicity are in units of $2\pi B_0^2 R^4$. The $2\pi$ factor accounts for the azimuthal integration. For example, a vacuum dipole has exactly $E_{\mathrm{dip}}=1/3$ for $B_1 = 1$.

The \texttt{python} script used to compute the solutions and produce the figures presented in this paper is made available on the dedicated Zenodo repository \footnote{\label{fn:zenodo} \url{https://doi.org/10.5281/zenodo.18155842}}. Unless otherwise stated, solutions to Eq. \eqref{eq:hierarchy} for the $F_i$ functions have been obtained numerically using the Runge-Kutta algorithm of order 4 with order 5 for error estimation (RK45) implemented in the \texttt{SciPy} library\footnote{\url{scipy.org}} \label{fn:scipy}

\subsection{General case with $c_{i\neq 2} = 0$} \label{sec:simpledip}
The most general solution can be expressed up to order $\bigcirc(1/r^5)$ as  
\begin{equation}
\label{eq:Br4}
\vB = \vB_{1}^{\rm v} + \vB_{2}^{\rm v, split} \pm  c_2 \frac{B_1^3}{8} \frac{(1-\mu^2)^{5/2}}{r^4} \ve_\varphi +\bigcirc\left(\frac{1}{r^5}\right),
\end{equation}
where $\vB_{1}^{\rm v}, \vB_{2}^{\rm v, split}$ are the vacuum dipole and split quadrupole, respectively, which we explicit in appendix \ref{ap:vacsol}, Eq. \eqref{apeq:B1v} and Eq. \eqref{apeq:B2split}. $B_1$ is the dipole strength at the pole. The dipole sources a toroidal field at order $1/r^4$ through Eq. \eqref{eq:A}, while both vacuum fields are purely poloidal. The $\pm$ sign in front of the toroidal term signifies the potential parity operation between hemispheres: the sign can be reversed across the equator for anti-twisted solutions. If so the solution generates a current sheet in the radial direction with surface density $\sigma_r = -2 c_2 B_1^3 (1-\mu^2)^{5/2}/r^4$, as obtained in Sec. \ref{sec:curdens}.

The twist of a field line is the difference of azimuth $\Delta\varphi = \int \mathrm{d}\varphi$ between its two footpoints integrated along the line.
In absence of split quadrupole, the twist of a magnetic field line emerging at colatitude $\theta$ is straightforward to calculate in the approximation of Eq. \eqref{eq:Br4},

\begin{eqnarray}
\label{eq:twist1}
\Delta \varphi(\theta) & = & 2\oint_{\theta}^{\pi/2} \frac{B_{\varphi}}{B_{\theta}} \frac{\mathrm{d} \theta'}{\sin\theta'} \\
\label{eq:twist2}
& = & \frac{1}{2} \frac{c_2 B_1^2}{R_*} \cos\theta \sin^2\theta, 
\end{eqnarray}
where in those units $R_*=1$.
The factor 2 in Eq. \eqref{eq:twist1} implies that this is the full twist from one hemisphere to the other. In the case of the anti-twisted configuration the hemisphere-to-hemisphere twist is by definition 0, but the twist between a pole and the current sheet is half the above value. 

Helicity is defined as $H = \int \vec{\Apot} \cdot \vec{B} \;\mathrm{d} V$ where $\vec{\Apot}$ is the magnetic potential vector. In the present case, one can show that (Appendix \ref{apsec:helicity})
\begin{equation}
	H = 2\int \frac{\ppot  B_\varphi}{r \sin\theta} \;\mathrm{d} V = \frac{4}{105} c_2 B_1^4.
\end{equation}
where the second equality is valid for the approximation of Eq. \eqref{eq:Br4}. As for the twist, this expression is valid only in the regularly twisted configuration, and $H=0$ in the anti-twisted case.

At order $1/r^5$, the equation for $F_3$ is sourced by $3c_2 G_5^{(5)} = F_1^5$, Eq. \eqref{eq:aA5}. As shown in appendix \ref{ap:F3}, this results in a solution with discontinuous $F_3'$ at the equator due to imposed even parity. This discontinuity is sustained by the toroidal component of the current sheet, Sec. \eqref{sec:curdens}, which therefore also decreases as $1/r^5$. Appendix \ref{ap:F3} also gives an analytical solution of the particular solution for $F_3$. Beyond this order, we consider in this paper that it is more economical to use numerical integration. Provided a quadrupole is also present, that is $F_2\neq 0$, a toroidal field will also be present at order $r^{-5}$ expressed by $\pm c_2F_1^2F_2/r^5\sin\theta$. More generally, none of the functions $F_{i\geq 3}$ are vacuum solutions, but instead are solutions of Eqs. \eqref{eq:hierarchy} sourced by combinations of functions of lower order. These higher-order functions, $F_{i\geq 3}$, are the ones that generate the equatorial discontinuity supported by the toroidal component of the current sheet. In the particular case where $F_2 = 0$ and  $\forall i > 1, \dot F_{2i} = 0$ one can show that all even-order functions are null, that is $\forall i > 1, F_{2i} = 0$.

\subsubsection{Dipole}
\label{sec:dipole}
\begin{figure}
	\centering
	\includegraphics[width=1\linewidth]{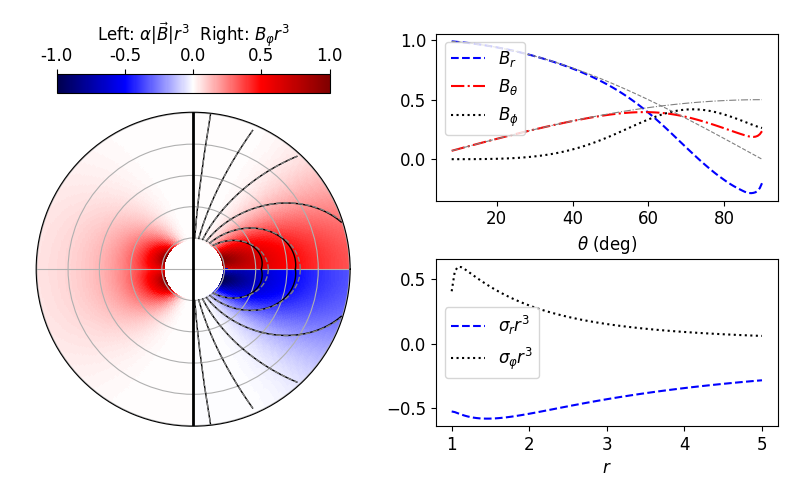} 
	\caption{Solution for $\dot F_1= B_1 =1, c_2=6$ between 1 and 5 stellar radii, where $B_1$ is the dipole field strength at the pole. All quantities are multiplied by $r^3$ for better visualisation. Left, left hemisphere: poloidal cross-section of the current amplitude $\alpha|\vec{B}| r^3$, normalised by its maximum. Left, right hemisphere: poloidal cross-section of the toroidal field $B_\varphi r^3$, normalised by its maximum. Poloidal magnetic field lines are shown as solid black lines, and vacuum-dipole field lines are shown for comparison as grey dashed lines.
		Top right: Magnetic-field components at the stellar surface, $B_r$ dashed line, $B_\theta$ dot-dashed line, and $B_\varphi$ dotted line in units of $B_1$. For comparison, vacuum-dipole components are plotted as grey lines with corresponding styles. Bottom right: current-sheet current components $\sigma_r$, dashed line, and $\sigma_\varphi$, dotted line, rescaled by a factor $r^3$.}
	\label{fig:B1c2}
\end{figure}

\begin{figure}
	\centering
	\includegraphics[width=0.49\linewidth]{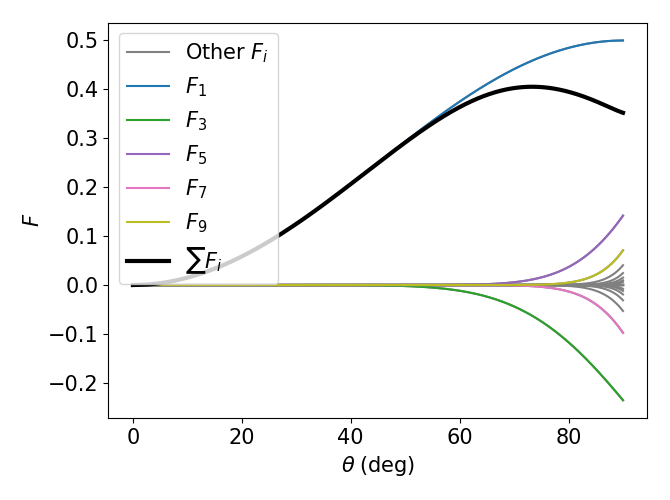} 
	\includegraphics[width=0.49\linewidth]{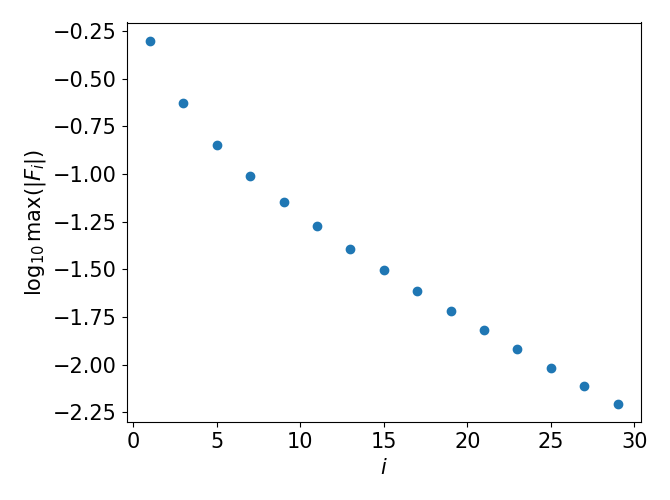} 
	\caption{Components (left) and convergence (right) of the solution to Eq. \eqref{eq:hierarchy} for the conditions of Fig. \ref{fig:B1c2}. Left: the components $F_i$ up to the truncation at order 30. For clarity only the lowest 9 orders are shown in colour and labelled, and the remaining ones are shown in grey. The thick black line represents their sum up to order 30 such that it represents the potential $\ppot$, Eq. \eqref{eq:ppot}, on the stellar surface at $r=1$.
	Right: convergence of the series defining $\ppot$, Eq. \eqref{eq:ppot}, for the conditions of Fig. \ref{fig:B1c2}. The maximum of each $|F_i(\mu)|$ for $\mu$ between 0 and 1 is plotted against its order $i$, where $F_i$ is solution of Eq. \eqref{eq:hierarchy}. Only odd orders are shown, as even orders are all equal to 0. }
	\label{fig:B1c2comp}
\end{figure}

In Fig. \ref{fig:B1c2} we show the anti-twisted solution sourced by a vacuum dipole. All multipoles have been computed up to order 30 and are shown in Fig. \ref{fig:B1c2comp}. Since no quadrupole is present, only odd orders $F_1, F_3, F_5...$ contribute. We can see in Fig. \ref{fig:B1c2} that at the surface the dipole component dominates up to a colatitude of $\sim 60^{\circ}$ and is dominated by higher order multipoles around the equator.
	Indeed, one can see in the top-right panel of Fig. \ref{fig:B1c2} that the radial magnetic component is far from following a dipolar behaviour as it flips sign within the northern hemisphere. This is primarily due to the influence of the octupolar component, $F_3$, at the surface, Fig. \ref{fig:B1c2comp}. However, since the latter component decays as $1/r^5$ it gives way to the dipolar component generated by $F_1$ after a few stellar radii. Thus, the solution is asymptotically dipolar. 

The largest twist angle is found to be $\Delta \varphi \simeq 1.20$ rad, close to the approximation of Eq. \eqref{eq:twist2}, $\Delta \varphi \simeq 1.15$ rad obtained at $\theta_{\max}\simeq 55$ deg. Similarly, magnetic helicity has $H \simeq = 0.16$, compared to the analytical value of 0.23. The magnetic energy is remarkably close to the vacuum dipole energy, indeed, $E/E_{\rm dip} \simeq 0.9995$ (numerically significant). About 12\% of the magnetic energy is stored in the toroidal component, mostly from its leading order contribution Eq. \eqref{eq:Br4}, which implies that the sourced higher multipoles partly cancel the dipolar component such that the total energy actually decreases slightly.

The value of $c_2 = 6$ is about the largest value that we found to converge, everything else being equal. The series of $F_i$ displays geometric convergence as can be seen in Fig. \ref{fig:B1c2comp}, as expected from the arguments given in appendix \ref{ap:convergence}.
We also show in appendix \ref{ap:convergence} that in this case the problem only depends on the parameter $\rho = 3(\dot F_1/2)^{4}c_2^2$ up to a scaling factor $\dot F/2$. This means that for a constant value of $\rho$,  $F_{i}/(\dot F_1/2)$ are identical.  

The components $F_{i > 1}$ in Fig. \ref{fig:B1c2comp} are all particular solutions, as a result of the choice of boundary conditions given by $\dot F_{i > 1}=0$. Indeed, a given solution is a linear combination of a vacuum and a particular solution, $F_i = F_i^{\rm v} + F_i^{\rm p}$. The power-series expansion of particular solutions are proportional to $\mubar^{l} = (1-\mu)^{l}$ where $l=i+2$ (see appendix \ref{ap:F3}). It follows that particular solutions have derivatives equal to zero at $\mu = 1$, that is $\dot F_i^{\rm p}{} = 0$. On the other hand, vacuum solutions all have a term $\bigcirc(\mubar)$ which means that the boundary condition $\dot F_i$ is entirely determined by the vacuum term. 

The higher the order of $F_i$ the more their effect is localised around the equator. This is a direct consequence of the fact that $F_i \propto \mubar^{i+2}$, which is visible in Fig. \ref{fig:B1c2comp}.


\subsubsection{Dipole + octupole}
\begin{figure}
	\centering
	\includegraphics[width=1\linewidth]{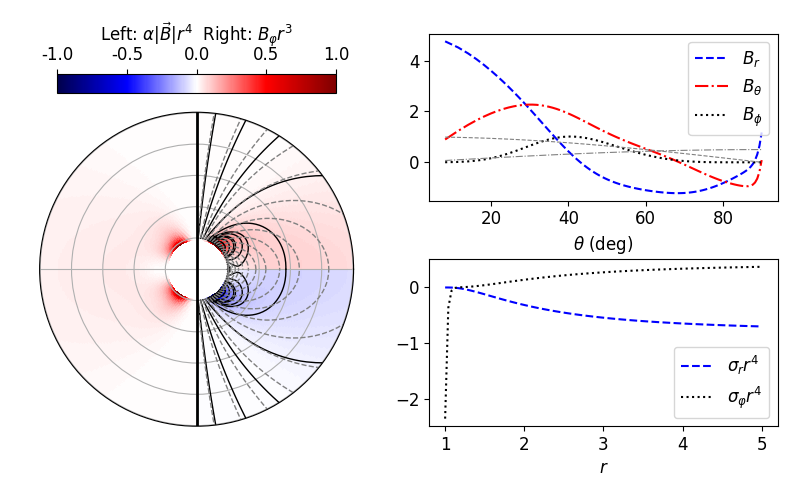} 
	\caption{Same as Fig. \ref{fig:B1c2} for $ \dot F_1 =1, \dot F_3 =4, c_2=3.2$ in the anti-twisted configuration.}
	\label{fig:b1b3c2}
\end{figure}

\begin{figure}
	\centering
	\includegraphics[width=0.49\linewidth]{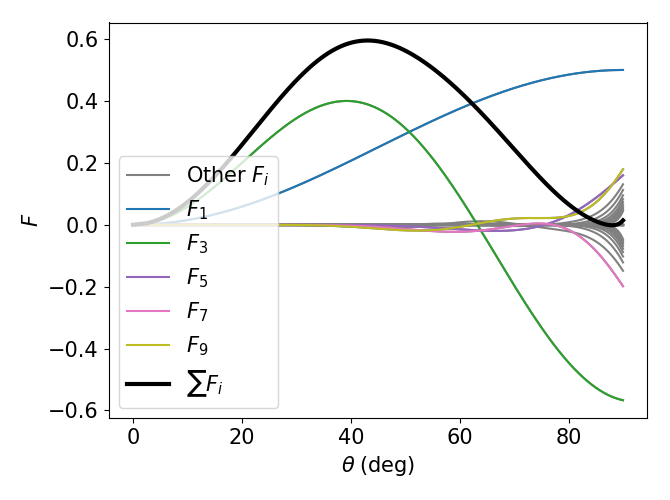} 
	\includegraphics[width=0.49\linewidth]{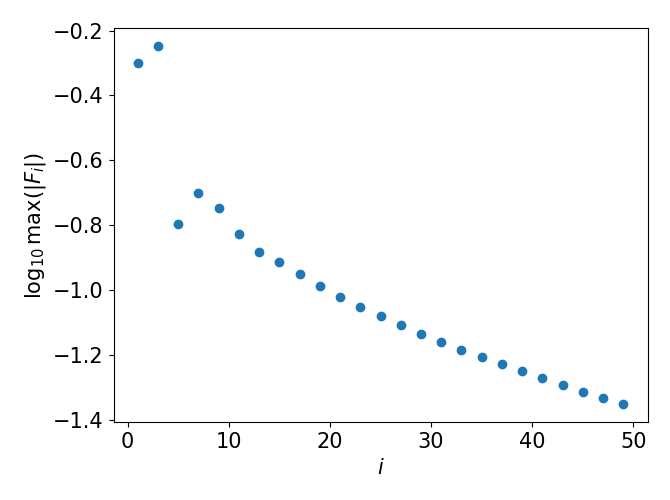} 
	\caption{Same as Fig. \ref{fig:B1c2comp} for the parameters of Fig. \ref{fig:b1b3c2}. The thick black (left panel) line shows the sum of the first 50 orders.}
	\label{fig:b1b3c2comp}
\end{figure}

In Fig. \ref{fig:b1b3c2}  we show the anti-twisted configuration of a vacuum dipole with the contribution of a vacuum octupole. We used the values $F_1' =-1, F_3' =-4 c_2=3.2$, where $c_2$ has about the largest value we found to allow for convergence in this case.  Figure \ref{fig:b1b3c2comp} shows the components separately, and one can clearly distinguish the octupole contribution, $F_3$, which slope at the origin is not null, meaning a vacuum contribution is present on top of the particular solution sourced by the dipole. The convergence plot in Fig. \ref{fig:b1b3c2comp} shows a monotonous convergence of $F_{i>5}$ while the lower orders are dominated by the effect of the values imposed on $F_1'$ and $F_3'$.

The result is a peak of the toroidal field at the surface at higher latitudes, around $\pm 45^{\circ}$. These contributions are however much more localised than in the dipole case above, as they result mainly from terms of the form $\propto c_2^2 F_3 F_1^2/r^6$. More generally, the slow convergence means that many multipoles contribute especially near the equator. This is particularly visible in the very steep decay of the toroidal component of the current sheet. Beyond a few stellar radii the structure of the field tends to that of Fig. \ref{fig:B1c2}. We note that, in this case as in the others, it is possible to produce a vanishing current sheet on the surface by fitting $\{\dot F_i\}$ so as to produce a continuous boundary condition, while the current sheet appears at higher altitude. This is particularly what is being done in Sec. \ref{sec:energytwist}.

The largest twist is here $\Delta \varphi \simeq 0.78$ rad, and helicity $H \simeq 0.18$. Similarly to the dipole case in Sec. \ref{sec:simpledip} the magnetic energy is very close to the vacuum energy (dipole + octupole) with $E/E_{\rm dip + oct} \simeq 0.991$ (numerically significant). About 4\% of the magnetic energy is stored in the toroidal component. However, the absolute amount of energy here represents $1.28 E_{\rm dip}$ due to the octupolar component. As a result, the absolute toroidal energy is larger than in Sec. \ref{sec:dipole} at about $0.14 E_{\rm dip}$. Although less twisted, this multipolar configuration represents a somewhat larger energy reservoir.

\subsection{Non-vacuum dipole + octupole}  \label{sec:nvdip}
The case $c_{i\neq 1} = 0$ has already been partially studied in \citet{low_modeling_1990}. Indeed, for $\dot F_{i\neq 1} = 0$, it is equivalent to the self-similar model discussed in Sec. \ref{sec:math} for index $p = 1$, and $F_1$ is the only function that is not null. There is a discrete spectrum of values of $c_1$ for which the function is fully continuous at the equator and does not require a current sheet. 
For $c_1 =0$ one has a vacuum dipole, and for higher values the angular dependency shows an increasing number of poles, but with a radial dependency that remains $r^{-3}$ for all components, including the toroidal one. The difference in this work, is that it is possible to add multipoles, for example by setting $\dot F_3 \neq 0$. This then triggers a cascade of higher order multipoles, similarly to the above cases. 

\begin{figure}
	\centering
	\includegraphics[width=1\linewidth]{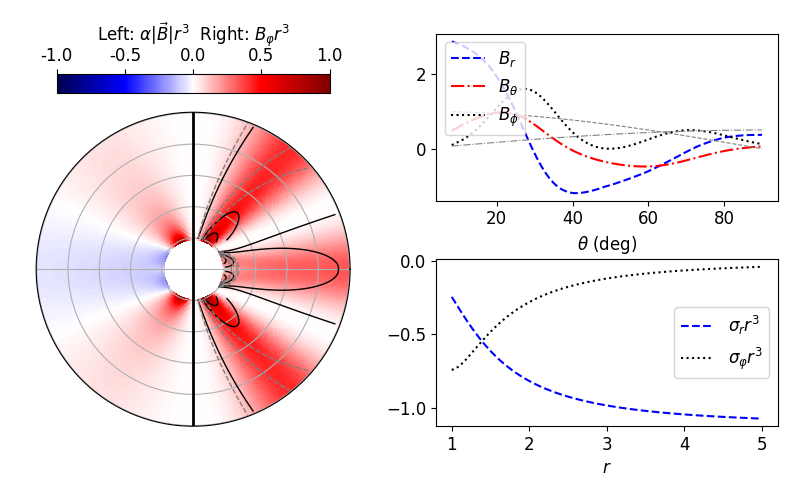} 
	\caption{Same as Fig. \ref{fig:B1c2} for $\dot F_1 =1, \dot F_3 =2, c_1=15.9$ and regular twist.}
	\label{fig:b1b3c1}
\end{figure}

\begin{figure}
	\centering
	\includegraphics[width=0.49\linewidth]{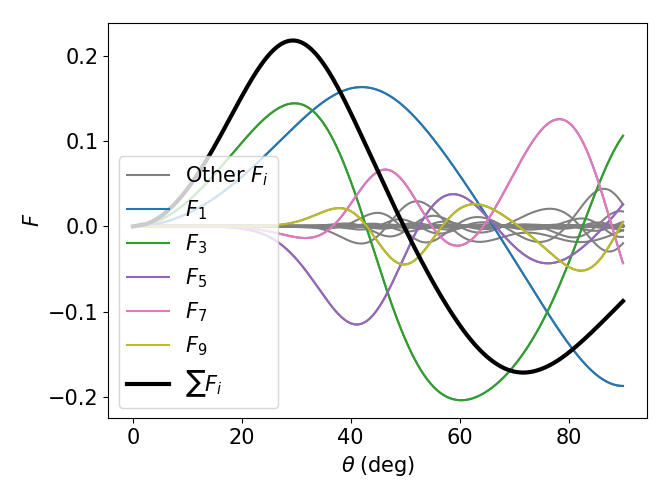} 
	\includegraphics[width=0.49\linewidth]{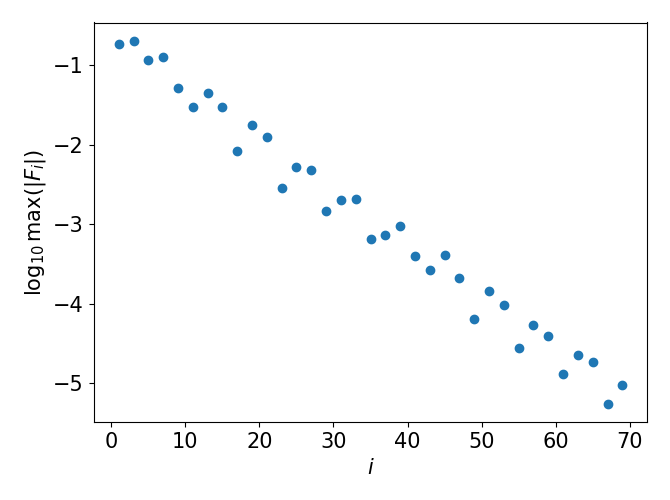} 
	\caption{Same as Fig. \ref{fig:B1c2comp} for the parameters of Fig. \ref{fig:b1b3c1}. The thick black (left panel) line shows the sum of the first 70 orders.}
	\label{fig:b1b3c1comp}
\end{figure}

This behaviour is illustrated in Figs. \ref{fig:b1b3c1}-\ref{fig:b1b3c1comp} where we have used $\dot F_1=1 ; \dot F_3=2$ at $\mu = 1$ and $c_1 = 15.9$, with every other initial conditions and constants equal to zero. The value of $c_1$ is determined by numerically solving $\dot{F}_1(\mu=0, c_1) =0$, which is the sufficient condition of $F_1$ to be even with respect to the equator. With this particular value of the constant $F_1$ is continuous at the equator and does not generate a current sheet. This is why here we show the twisted case instead of the anti-twisted one. A current sheet with only a toroidal component is however still supported by the higher orders, mainly $F_3$.

The maximum twist in this configuration reaches $\Delta \varphi \simeq 5.8$ rad. This extreme value is related to the homogeneous radial dependency of all components much more than to the octupolar component. Indeed, it is very close to the value obtained without the additional octupole. Helicity is $H \simeq 0.088$. 

The magnetic energy compared to the vacuum case (dipole + octupole) is $E/E_{\rm dip+oct} \simeq 1.05$. Consistently with the large twist, the toroidal energy is proportionally larger and represents about 27\% of the total magnetic energy. Contrary to the two other cases, the energy is larger than in the vacuum case, and the difference between the two is also more significant. We conjecture that this is to put on the account of the non-linear nature of this configuration where, contrary to the two previous cases, the top-level dipole is itself modified compared to vacuum.


\subsection{Surface boundary conditions, energy, and twist}
\label{sec:energytwist}
\newcommand{\dpot}{{\cal D}}
We have so far considered solutions where the value of $\ppot$ on the surface of the star was a by-product of the solution rather than a boundary condition. This is because the boundary conditions of the problem are conveyed by the constants $\{\dot F_i\}$ which are not straightforwardly connected to the surface potential. On the other hand, given that the inner boundary condition is determined by an infinite number of degrees of freedom, $\{\dot F_i\}$, we can conjecture that this class of solutions allows for arbitrary conditions at the stellar surface. This is not, for example, the case of self-similar solution in Sec. \ref{sec:math}.

Here, we take the opposite route: the surface potential is fixed to the vacuum dipole value, i.e.  $\ppot(r=1, \mu) =  \dpot(\mu)$ where $\dpot=F_1^{\rm v}$ is analytically defined by Eq. \eqref{apeq:vacdipole}. As before we set $B_1=\dot F_1^{\rm v}=1$ such that $\dpot \sim 1$.  To do so we use a Nelder-Mead minimizer (see footnote \ref{fn:scipy}) in order to fit the $\{\dot F_i\}$ such that the boundary condition is fulfilled. As a metric, we use $\chi^2 = \sum_j \left(\ppot(r=1, \mu_j\right) - \dpot(\mu_j))^2$ where we take 100 evenly spaced $\mu_j$ to sample the function. The task is potentially difficult due to the large number of parameters. Fortunately we found that a limited number of $\dot F_i$ is sufficient for a reasonable accuracy. In practice, for $c_2 < 4$ we limited ourselves to twenty multipolar orders, but fitted only $\dot F_{i\leq 10}$ while $\dot F_{i> 10}=0$. For $c_2 \geq 4$, we used 28 orders and fitted the first 14. This is because as the coupling constant increases the source terms remain significant at higher orders. The absolute difference with the target boundary condition is $\max(\|\ppot - \dpot\|) \leq 0.0001$.

\begin{figure}
	\centering
	\includegraphics[width=0.7\linewidth]{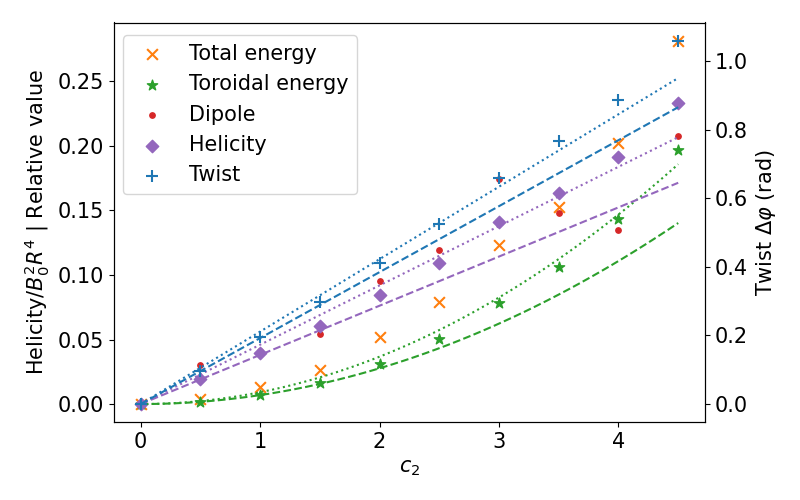}
	\caption{Left axis: relative total magnetic energy $(E - E_{\rm dip})/E_{\rm dip}$ (orange "x" markers), relative toroidal magnetic energy $E_{\rm toro}/E_{\rm dip}$ (green stars), dipolar component strength with respect to a vacuum dipole $\dot F_1/\dot F_1^{\rm v}$ (red circles), and helicity $H$ in units of $B_0^2R^4$ (purple diamonds). Right axis: maximum twist angle (blue "+" markers). 
		All quantities are computed for a sequence of solutions characterised by the value of $c_2$ and with $c_{i\neq2}=0$. These solutions share the same boundary condition $\ppot(r=1)=\dpot$ at the stellar surface. Dashed lines represent the twist (blue), helicity (purple), and relative toroidal energy (green) computed using the analytical approximation in Eq. \eqref{eq:Br4} with the correct boundary conditions  $\dot F_1 = \dot F_1^{\rm v}$. Dotted lines correspond to a slightly stronger dipole  $\dot F_1 = 1.048\dot F_1^{\rm v}$.}
	\label{fig:degsol}
\end{figure}

The results are shown in Fig. \ref{fig:degsol}, and compared to the analytic approximation of Eq. \eqref{eq:Br4}. The coupling constant $c_2$ ranges in $[0, 4.5]$. For larger values we found that the solutions do not converge with this boundary condition. We found that the maximum twist angle and helicity are nearly linear functions of $c_2$ and the toroidal energy a quadratic function. This is in agreement with the analytic approximation, which is particularly good for smaller values of $c_2$. As seen in other solutions \citep{akgun_force-free_2016}, the dipolar component increases with increasing currents and as such is largely responsible for the increase in total magnetic energy. Indeed, we can see that the total energy increases faster than the toroidal component, while the increase is exactly the same in the analytical model, since the dipole remains constant and there are no additional multipole. In the exact solution, multipolar orders also increasingly contribute to the energy budget. Finally, we stress that although the dipolar component increases by $\sim 20\%$ compared to vacuum, the surface field is nonetheless that prescribed by the boundary condition, where other components effectively screen the enhanced dipole. 

We performed a fit of the toroidal energy with the analytical model in Eq. \eqref{eq:Br4} in order to find the effective dipolar amplitude $B_1$ that allows to best reproduce the data. We found $B_1 \simeq 1.048$ to work best, and it allows to reproduce quite accurately not only the toroidal energy but also helicity as well as the maximum twist. Thus, we may say that as far as these quantities are concerned, the exact solution effectively behaves as the analytical approximation for a surface boundary condition given by a vacuum dipole about 5\% stronger.


\section{Discussion and conclusion} \label{sec:conclusion}
We present in this work a class of solutions to the Grad-Shafranov equation based on a multipolar expansion the angular part of which is determined by a hierarchy of ordinary differential equations as exhibited in Eq. \eqref{eq:hierarchy}. Solutions are characterised by a set of arbitrarily chosen coupling constants $\{c_i\}_{i\geq 1}$ and initial conditions for the derivatives of the angular functions $\{\dot F_i\}_{i\geq 1}$. The main limitation in the choice of couplings is the necessary convergence of the series. The multipolar vacuum expansion is recovered when all coupling constants are zero.

In Sec. \ref{sec:application} we present a subset of solutions symmetric about the equatorial plane and supported by an equatorial current sheet (in particular Figs. \ref{fig:B1c2}, \ref{fig:b1b3c2}, \ref{fig:b1b3c1}). The twist, or equivalently the toroidal component of the magnetic field, can have opposite signs in opposite hemispheres. However, the physical interpretation of such anti-twisted configurations is unclear.

Current sheets have been previously obtained (see in particular the numerical work of \citet{parfrey_dynamics_2013} and references therein), and are expected to form past a certain level of twisting. Here, however, the current sheet is part of the stationary solution, albeit presumably dynamically unstable to reconnection. The case of reversed toroidal fields across the equator is reminiscing of the formation of similar internal magnetar configurations due to the Hall drift \citep{vigano_new_2012}. 

A case of interest is when coupling constants are chosen such that the magnetosphere asymptotically tends to vacuum. In such case, the hierarchy of equations is linear, and can possibly be solved through power-series expansions. However the complexity of the source terms makes it rather unwieldy and one may prefer a numerical solution. To this effect, we share the \texttt{python} script used for this work (see footnote \ref{fn:zenodo}).

We show in Eq. \eqref{eq:Br4} the simplest solution truncated at order $r^{-5}$ that asymptotically tends to a vacuum dipole. This results in a simple analytical expression where the leading-order toroidal components intervenes at order $r^{-4}$ and is stronger near the equator. This approximation is valid at distance sufficiently large to neglect multipolar-order contributions, but is also a good approximation for low to moderate values of the coupling constants of solutions with dipolar inner boundary condition, reproducing well the twist and toroidal magnetic energy, in particular (Sec. \ref{sec:energytwist}).
 
The class of solutions described in this paper has an infinity of degrees of freedom, both in terms of coupling constants $c_i$ and initial condition $\dot F_i$ for the angular functions. Both can in principle be adjusted in order to produce a particular surface field, for example to match an internal solution. The question of whether this freedom allows one to reproduce any desired surface condition, that is the multipolar expansion forms a complete set of functions, is open. In addition, this method can become impractical if too many components are needed. So far, our attempts at fitting dipolar boundary conditions in Sec. \ref{sec:energytwist} was quite straightforward, involving only a limited number of parameters.

Thus, we could study the evolution of the magnetosphere as a function of current coupling for fixed boundary conditions. As in other works \citep[e.g.]{akgun_force-free_2016} we find a $\sim 25\%$ increase of the dipolar component with increasing currents, as well as enhanced multipolar orders, leading to a similar increase in total energy. Toroidal energy increases less, $\sim 15-20\%$ reflecting the amplification of poloidal fields. The maximum twist increases quasi-linearly and all our solutions failed to converge for twists little above 1.1 rad. Overall, we could not obtain twists larger than $\sim 1.2$ rad for asymptotically dipolar configurations. This is similar to critical values found in previous numerical work on stationary solutions \citep[][and references therein]{akgun_force-free_2016} which typically to not exceed $\sim 1.5$ rad, but lower than dynamical studies which reach values of $\sim 3$ rad \citep{parfrey_dynamics_2013, mahlmann_three-dimensional_2023}.  As an exception, the strongly non-linear solution presented in Sec. \ref{sec:nvdip} reaches a twist of 5.3 rad, but this solution is not asymptotically dipolar.

\section*{Supplementary data}
The scripts implementing the solution described in this paper and generating all the figures are available at \url{https://doi.org/10.5281/zenodo.18155842}

\section*{Acknowledgements}
The author thanks Dr. Fabrice Mottez (Observatoire de Paris, CNRS, France) for his advice and support, and Dr Kōji Uryū (University of the Ryukyus, Okinawa, Japan) for helpful discussions about his 2023 paper. The author would like to thank the anonymous referees for their constructive comments that helped improving this work. 

\section*{Funding}
This research received no specific grant from any funding agency, commercial or not-for-profit sectors.

\section*{Declaration of interests}
The author report no conflict of interest.

\section*{Author ORCID}
G. Voisin, \url{https://orcid.org/0000-0003-2845-8905}

\bibliographystyle{jpp}
\bibliography{FFtwist} 

\makeatletter
\def\fps@table{h}
\def\fps@figure{h}
\makeatother

\section*{Appendix}

\section*{Vacuum solutions} \label{ap:vacsol}

Vacuum solutions to the Grad-Shafranov equation are usually expressed using Legendre polynomials \citep[see  e.g.][]{vigano_force-free_2011}. For the purpose of this work we find it insightful to outline a demonstration of these solutions and explicit the first multipoles. 

In vacuum Eq. \eqref{eq:hierarchy} becomes,
\begin{equation}
\label{apeq:vacuumeq}
\forall i \geq 1, -l(l+1) F_l^{\rm v} - (1-\mu^2) {F_l^{\rm v}}'' = 0. 
\end{equation}
We remark that this could directly have been obtained from the Grad-Shafranov equation, Eq. \eqref{eq:gseq}, with an ansatz of the form $\ppot = F_l^{\rm v}(\mu)/r^l$. 

Since we specify boundary conditions at $\mu = 1$, we will seek a power-series solution of the form $F_l^{\rm v} = \sum_{i=0} a_i \mubar^i$, where $\mubar = 1- \mu$. Indeed, expansions around $\mu=0$ do not always converge at $\mu =1$. Injecting this series into \eqref{apeq:vacuumeq}, one finds after a few manipulations that solutions to the equations follow the sequence defined by
\begin{equation}
\label{apeq:vacseq}
\forall i>0, a_{i+1} = a_i \frac{i(i-1) - l(l+1)}{2i(i+1)},
\end{equation}
and $a_0 = 0$. 
This sequence is initialised by choosing $a_1$. One sees that $a_{i > l+1} = 0$, such that these solutions are finite polynomials. Since $a_0 =0$, these solutions satisfy $F_l^{\rm v}(\mubar = 0) = 0$, as required. It is less obvious what their behaviour is at $\mubar = 2$ ($\mu = -1$). A change of variable $\mubar \rightarrow 1-\mu$ shows that these solutions are even for odd values of $l$ and odd for even values, meaning that $F_l^{\rm v}(\mu = \pm 1) = 0$, and all required boundary conditions are matched. 
It is immediate to see that $a_1 = -{F^{\rm v}}'(\mu=1) = \dot F^{\rm v}(\mubar=0)$ where the dot denotes differentiation with respect to $\mubar$. 

These solutions form a multipole basis. We explicit the first orders, 
\begin{eqnarray}
\label{apeq:vacdipole}
F_1^{\rm v} & = & a_1 (\mubar - \frac{1}{2}\mubar^2) = \frac{a_1}{2}(1-\mu^2), \\
F_2^{\rm v} & = & a_1\mubar\left(1 - \frac{3}{2}\mubar +\frac{1}{2}\mubar^2\right) = \frac{a_1}{2}\mu(1-\mu^2) , \\
\label{apeq:f3}
F_3^{\rm v} & = & a_1\mubar \left(1 - 3\mubar +\frac{5}{2}\mubar^2 - \frac{5}{8}\mubar^3\right)  \\ 
& = & -\frac{a_1}{8}(1-6\mu^2 + 5\mu^4), \nonumber
\end{eqnarray}

Using Eq. \eqref{eq:Bp} with $\ppot = F_l/r^l$, we obtain expressions for the vacuum magnetic fields, 
\begin{eqnarray}
\label{apeq:B1v}
\vB_{1}^{\rm v} & = &  \frac{B_1}{r^3}\left(\cos\theta \ve_r + \frac{\sin\theta}{2}\ve_\theta \right), \\
\label{apeq:B2v}
\vB_{2}^{\rm v} & = & \frac{B_2}{r^4} \left((2\cos\theta - 1)\ve_r - \cos\theta\sin\theta \ve_\theta\right), \\ 
\label{apeq:B3v}
\vB_{3}^{\rm v} & = &  \frac{B_3}{r^5} \left((\frac{5}{2}\cos^4\theta - \frac{3}{2}\cos\theta)\ve_r + (\frac{\sin^3}{8}\theta  - \frac{\sin\theta}{2})\ve_\theta \right),
\end{eqnarray}
where $B_1, B_2, B_3$ are the magnetic strengths at the pole corresponding respectively to $a_1, a_1/2, a_1$ in the notations of Eqs. \eqref{apeq:vacdipole}-\eqref{apeq:f3} 

In the case of imposed even parity across the equator, the quadrupole field becomes a split quadrupole which can be written 
\begin{equation}
\label{apeq:B2split}
\vB_{2}^{\rm v, split} = \left(H(\mu) - H(-\mu)\right) \vB_{2}^{\rm v} ,
\end{equation}
where $H(x)$ is the Heaviside step function such that $H(x\geq 0) = 1$ and $H(x<0)=0$.

\section*{Example of analytical solution for $F_3$}\label{ap:F3}
\newcommand{\bara}{\bar a}

We consider the case where $F_1$ is a vacuum dipole, Eq. \eqref{apeq:vacdipole}, $ F_2 = 0$ and $c_2$ is the only non-zero coupling constant.
In this case, the source term of the $F_3$ equation in Eq. \eqref{eq:gseq} is $ [\alpha A]_5 = 3c_2^2 G_5^{(5)} = 3c_2^2  F_1^5$ as can be seen from Eq. \eqref{eq:aA5}. 

We solve the equation for $\phi_3 = F_3/(3c_2^2 \dot F_1(0)^5)$ where we denote $\dot F $ the derivative of $F$ with respect to $\mubar = 1-\mu$,
\begin{equation}
\label{apeq:F3}
-12  \phi_3 - \mubar(2-\mubar) \ddot \phi_3 = \left(\mubar - \frac{1}{2}\mubar^2\right)^5.
\end{equation}

In order to get a particular solution $\phi_3^{\rm(p)}$ we proceed exactly as for the vacuum solutions of Sec. \ref{ap:vacsol}, by expanding $ \phi_3^{\rm (p)} = \sum_{i=0} \alpha_i \mubar^i$, obtaining 
\begin{equation}
\sum_{i=0}\mubar^i \left[-2(i-1)i \alpha_{i+1} - \alpha_i \left(l(l+1) - i (i-1)\right)\right] = \left(\mubar - \frac{1}{2}\mubar^2\right)^5,
\end{equation}
where here $l = 3$.

We find the particular solution 
\begin{eqnarray}
\label{apeq:phi3}
\phi_3^{\rm (p)} & = & -\frac{8}{15} \mubar^6 + \frac{88}{105}\mubar^7 - \frac{24}{49}\mubar^8 + \frac{113}{882} \mubar^9 - \frac{17}{1323}\mubar^{10} \\
&  & - \frac{1}{97020}\mubar^{11} + \sum_{i=12}^{\infty}\alpha_i \mubar^i \nonumber
\end{eqnarray}
where the polynomial part results from matching the first 11 terms with the source term on the right-hand side of Eq. \eqref{apeq:F3}, and the series is generated by using the relation in Eq. \eqref{apeq:vacseq} (replacing $a$ with $\alpha$) for $i \geq 11$ seeded with $\alpha_{11} = 1/97020$.
The terms up to order $\bigcirc(\mubar^5)$ do not appear here because they correspond to the vacuum solution in Eq. \eqref{apeq:F3}. The full solution is thus $F_3 =  F_3^{\rm v} + 3c_2^2 \dot F_1(0)^5 \phi_3^{\rm (p)}$ where the first term is given by Eq. \eqref{apeq:F3} and the second one by Eq. \eqref{apeq:phi3}. We have checked that Eq. \eqref{apeq:phi3} matches the numerical integration used in the main text within numerical error.

We see that the particular solution of Eq. \eqref{apeq:phi3} fulfils the boundary condition $\phi_3^{\rm (p)}(\mubar=0) = 0$ and that its first derivative is equally null $\dot \phi_3^{\rm (p)}(\mubar=0) = 0$, meaning that the slope at the origin is entirely given by the vacuum solution. More generally, it can be shown by induction that the source terms of the equation of order $i$ are proportional to $ \mubar^{l}(1 + \bigcirc(\mubar))$, with $l= i+2$, implying that $\phi_i^{\rm (p)} = \bigcirc(\mubar^l)$.

On the other hand, this solution is neither symmetric with respect to $\mu = 0$ nor does it fulfil the second boundary condition $\phi_3^{\rm (p)}(\mubar=2) = 0$. Indeed, we find numerically $\phi_3^{\rm (p)}(\mubar=2) \simeq 0.057$. As a consequence, in order to respect the boundary conditions we construct a solution by forcing a symmetry with respect to the equator. In particular, this implies a discontinuity of $\dot F_3$ at the equator, which results in a discontinuity of the radial magnetic field supported by a toroidal current sheet (see main text). 

The series term in Eq. \eqref{apeq:phi3} converges for $\mubar \leq 2$, however its derivative is not defined at $\mubar = 2$ as its series diverges. Indeed one can show that at $\mubar = 2$ the series is asymptotically given by $\alpha_i \mubar^i \sim 1/i^2$, the derivative of which is $\sim 1/i$ and thus diverges. From a practical point of view, we note that the polynomial part of Eq. \eqref{apeq:phi3} is accurate to $\sim 10^{-5}$, which we expect to be sufficient for most applications. 

\section*{Convergence}\label{ap:convergence}
\newcommand{\cL}{{\cal{L}}}

\subsection{Discussion of the $c_{i\neq 2}$ case}
A case of interest that we exhibit in Sec. \ref{sec:application} is when $c_{i\neq 2} = 0$, $F_1 \neq 0$, and $F_{i>1} =0$. That is, we have a vacuum dipole which sources higher multipoles through the coupling $c_2$. Only functions $\{F_{2i+1}\}_{i\geq 1}$ are non zero, and the fact that $F_{2i+1}'=0$ means that all solutions are particular solutions without vacuum terms (see main text and appendix \ref{ap:F3}). We show below that the right-hand side of Eq. \eqref{eq:hierarchy} is, up to a factor of order 1, $[\alpha A ]_{l=i+2} \propto \beta_i = \rho^i F_1'/2$ for $i \geq 3$, where $\rho = 3(F_1'/2)^{4}c_2^2$. In addition, we show that dividing both sides of each equation in Eq. \eqref{eq:hierarchy} by $F_1'/2$ results in $F_i/F_1'$ depending only on $\rho$.

We call $\chi_i = F_i/(F_i'/2)$ and get for the first orders, 
\begin{eqnarray}
\cL_1 \chi_1 & = & 0 \\
\cL_3 \chi_3 & = & \rho \chi_1^5 \\ 
\cL_5 \chi_5 & = & \rho n_1 \chi_3 \chi_1^4 = \bigcirc\left( \rho^2 \right) \\
\cL_7 \chi_7 & = & \rho (n_2\chi_5 \chi_1^4 + n_3\chi_3^2\chi_1^3) = \bigcirc\left( \rho^3 \right) 
\end{eqnarray}
where $\cL_i = -i(i+1)  - (1-\mu^2) \partial_{\mu^2}^2 $ is the linear operator on the left-hand side of Eq. \eqref{eq:hierarchy}. The $n_i$ are the number of terms involved that can be calculated from Eq. \eqref{eq:g}, but they are irrelevant for our current discussion. 

By definition $\chi_1 \sim 1$. Indeed it is a vacuum solution, Eq. \eqref{apeq:vacdipole}, such that $\max(|\chi_1|) =1$. It follows that $\chi_3 \sim \rho$, $\chi_5 \sim \rho^2$ up to a factor of order 1. The fact that particular solutions of $\chi_i$ are proportional to the amplitude of the right-hand side is immediately apparent by looking for series solutions as we do in appendix \ref{ap:F3}.

We proceed by induction to show that $\chi_{2i +1} \sim \rho^i$. Initialisation has just been shown, therefore we assume that this is true for all functions up to order $2i +1$. At order $2i+3 $, we see from Eq. \eqref{eq:g} that the right-hand side is 
\begin{equation}
\rho G_{l+2}^{(5)} =  \sum_{i_1+...+i_5 = l+2; i_j \geq 1} \chi_{i_1}...\chi_{i_5},
\end{equation}
with $l =2i+1 +2$. 
We see that the only combinations of ${i_1, i_2, i_3, i_4, i_5}$ possible can be obtained by taking those present at order $l$ in $G_{l}^{(5)}$ and by replacing a single $i_j$ by $i_j+2$. In these terms the largest $i_j$ is $2i-1$ for the term $\chi_{2i-1} \chi_1^4$. Since $\chi_{i_j +2 } \sim \rho \chi_{i_j}$, these terms are indeed of order $\bigcirc(\rho^{i+1})$. Thus the induction is proven. 

Based on this argument, we expect that the $\chi_{2i+1}$ can be bounded by a geometric sequence with common ratio $\sim \rho$. Convergence then corresponds to $\rho \lesssim 1$, which means $|c_2| \lesssim 2.3 F_1'(1)^2$. Numerically, we observe approximate geometric convergence up to $|c_2| \lesssim 6 F_1'(1)^2$. It is to be noted, that we have here neglected the increasing number of terms on the right-hand side. However one can see by looking at Eq. \eqref{eq:g} that, without detailed calculation, this number can be bounded by $l^4$. Thus the number of terms increases polynomially, which does not affect geometric convergence. 

The series that actually must converge is $\chi_i/r^l$. It follows that if $\chi_{2i+1} = \bigcirc(\rho^i)$ then  $\chi_i/r^l= \bigcirc\left((\rho/r)^i\right)$.. As a result, if the series converges at the surface, for $r=1$, then it converges for any $r \geq 1$. If $\rho = \rho_{\max}$, the largest value for which convergence is allowed at $r=1$, then the series is expected to diverge for $r < 1$. This is the behaviour we numerically observe. 

%
%

\section*{Parity}
\label{apsec:parity}
We want to show that the source terms of even-order functions $F_{2i}$ are odd and even for odd-order functions $F_{2i-1}$ in the case $c_{i\neq 2} =0$.
The source term of the equation for $F_i$ is $3c_2^2 G_{i+2}^{(5)}$ for $i \geq 3$ as can be seen from Eq. \eqref{eq:alphaA}. For $i = 3$, $G_{5}^{(5)} = F_1^5$ is even, as desired. For $i=4$, $G_{6}^{(5)} \propto F_2 F_1^4$, which is odd, equally as desired. Equation \eqref{eq:g} shows that $G_{l}^{(5)} = \sum_{i_1, .., i_5} F_{i_1}..F_{i_5}$ with $i_1+..+i_5=l$ and $l=i+2$. From Eq. \eqref{eq:gdep}, we know that $G_{l}^{(5)}$ depends only of function $F_{j<i}$, such that we can prove by induction the result. Given $i>2$, we assume that parity properties are fulfilled for all $j<i$. Then, for any $l > 4$, if $l$ is even then an even number of odd indices is required which means that the number of even indices is odd. It follows that $G_{l}^{(5)}$ is odd for even $l$, as desired. Conversely, for odd $l$ we need an odd number of odd indices and an even number of even indices, such that $G_{l}^{(5)}$ is even, as desired. Thus, we have proven the result.

\section*{Magnetic helicity} \label{apsec:helicity}
One can check that 
\begin{equation}
	\vec{\Apot} = \left(-r\int_{0}^{\theta}B_\varphi\mathrm{d}\theta',0,\frac{\ppot}{r\sin\theta}\right)_{(r,\theta, \varphi)}
\end{equation}
forms a suitable vector potential for the solution presented in Sec. \ref{sec:solution}.

Helicity is defined by $H = \int \vec{\Apot} \cdot \vec{B} \;\mathrm{d} V$. The radial term of the scalar product gives 
\begin{equation}
	-\int \frac{\partial_\theta \ppot }{r^2 \sin\theta} r\int_{0}^{\theta}B_\varphi\mathrm{d}\theta' \;\mathrm{d} V = \int \frac{\ppot  B_\varphi}{r \sin\theta} \;\mathrm{d} V
\end{equation}
where the right-hand side is obtained by integration by part in $\theta$. 
Adding up the other term of the scalar product we obtain,
\begin{equation}
	H = 2\int \frac{\ppot B_\varphi }{r \sin\theta}\;  \mathrm{d} V.
\end{equation}

\end{document}